\documentstyle[12pt,a4,psfig,here]{article}
\begin{document}
\title{ Rare Decay of K Meson 
\footnote{Talk presented by Y. Kiyo at Workshop on ``Fermion Mass and CP
Violation'', Hiroshima, 5-6 March 1998. 
} }
\author{ {\normalsize\sc Y. Kiyo, T. Morozumi }\\
\normalsize\sl Dept. of Phys. Hiroshima University 1-3-1 Kagami-yama \\
\normalsize\sl Higashi-Hiroshima 739-8526, Japan \\
{\normalsize\sc M. Tanimoto}\\
\normalsize\sl 
\normalsize\sl Science Education Laboratory, Ehime University\\
\normalsize\sl Bunkyo-cho, Matsuyama 790-8577, Japan}
\date{}
\maketitle
\begin{picture}(5,2)
\put(380,220){HUPD-9809}
\put(380,210){hep-ph/9805307}
\end{picture}
%------------- abstract -------------------------
\begin{abstract}
{\normalsize
We investigate the rare decay processes of the K mesons, 
$K_{L,S} \rightarrow \pi^{0} \nu \bar{\nu}$ and 
$K^{+} \rightarrow \pi^{+} \nu \bar{\nu}$ in  LR model 
with lepton family number being well conserved. 
In these processes, scalar operators $(\bar{s}d)(\bar{\nu}_{\tau}\nu_{\tau})$,
which are derived from 
 box diagram in LR model, play an important role due to 
an enhancement factor $M_{K}/m_{s}$ in the matrix element 
$<\pi|\bar{s}d|K>$. It is  emphasized   that the $K_{L}$ 
decay process through the scalar operator is not the CP violating 
mode, so the $B(K_L \rightarrow \pi^{0}\nu \bar{\nu})$ remains 
non-zero even in the CP conserved limit. We present  the pion energy spectrum 
for these processes and discuss the effects of  LR model.
}
\end{abstract}

%------------- introduction -------------------------
\section{Introduction} 
Decay of the neutral K meson taught us the violation of CP symmetry.
Our understanding about the CP violation is based on one complex phase
in the Cabibbo-Kobayashi-Maskawa (CKM) matrix \cite{CKM}, 
however, our knowledge of the CKM phase  is poor today. 
The projects of the B-factory are starting at KEK and SLAC,  
where  CKM sector of SM will be tested by using  e.g. unitarity triangle in the B 
meson system. The precise determination of the CKM parameters will be one of the
most important progresses to understand the nature, physics of 
violated symmetry. 

Experiments in the K meson system have entered new period.
That is the observation of the rare process, $K^+ \rightarrow
\pi^{+}\nu \bar{\nu}$ and the search for $K_{L} \rightarrow \pi^{0} \nu \bar{\nu}$. 
Recently, the signature of decay $K^+ \rightarrow \pi^+  \nu
\bar{\nu}$ has been observed by E787 Collaboration \cite{E787} 
and the reported branching ratio is $4.2^{+9.7}_{-3.5}\times 10^{-10}$, 
which is consistent with the expectation value in SM. 
Additional data are expected as well as the  improvement of the experimental data
 in the near future \cite{KOMA}. This situation forces  us the detail study of the 
rare decays of 
the K mesons. The decay $K_L \rightarrow \pi^0 \nu \bar{\nu}$ is one of the most
promissing process since it is  a CP violating mode in SM. 
This mode is theoretically clean to extract the CKM 
parameter $\eta$ \cite{BURA}.  

In this paper, we investigate rare decays of the K mesons 
in Left-Right (LR) model \cite{MORO} introducing right handed 
neutrinos. However, in the model, the neutrino masses are zero in
the tree level and lepton flavor is well conserved \cite{PREP}
(Also see analyses in the other models \cite{SEE}).
There appear scalar and tensor operators 
$(\bar{s}\left[1,\sigma^{\mu\nu}\right]d)
(\bar{\nu}\left[1,\sigma_{\mu\nu}\right]\nu)$ from the LR box
diagrams, in which left and right handed gauge bosons $W_{L}$ and 
$W_{R}$ are exchanged. 
The scalar operators have an enhancement factor $M_{K}/m_{s}$ 
in the matrix element $<\pi|\bar{s}d|K>$
( $M_K$ and $m_s$ is K meson and strange quark mass respectively). 
Thus the scalar operator may have large contribution to the rare 
decays of K mesons, $K^+ \rightarrow \pi^+ \nu \bar{\nu}$ 
and $K_{L,S} \rightarrow \pi^0 \nu \bar{\nu}$. 
An important point is that the CP property of the scalar interaction 
is different from the V-A interaction 
$(\bar{s}d)_{V-A}(\bar{\nu}_{l}\nu_{l})_{V-A}$ in SM.
The decay $K_L \rightarrow \pi^0 \nu \bar{\nu}$ through 
the scalar operator is not CP violating one, so we have non-zero 
branching ratio $B(K_L \rightarrow \pi^0 \nu_{l} \bar{\nu}_{l})$ even in 
the CP conserved limit ($\eta \rightarrow 0$).
Thus, it is interesting to estimate how large the effect 
of the scalar operator in the  pion energy spectrum. 

This paper is organized as follows. In section 2, we first 
discuss the rare decays of the K meson rather generally with 
an effective Lagrangian including the scalar interaction.
The scalar interaction in LR model is briefly discussed.
In section 3, we show the pion energy spectrum in LR model,
section 5 is devoted to discussion and summary.

%------------- Sec.2 -------------------------
\section{Rare Decay Through the Scalar Operator}
Rare decays, $K^+ \rightarrow \pi^+ \nu \bar{\nu}$ and 
$K_{L,S} \rightarrow \pi^0 \nu \bar{\nu}$, are flavor changing 
neutral current (FCNC) processes which are loop-induced one in SM and 
short-distance dominated one, so theoretically very clean modes 
\cite{BURA}.
The matrix elements involved in these  decays are related to 
experimentally well known leading decay $K^+ \rightarrow \pi^0 e^+ \nu$
using isospin symmetry, and  corrections to this relations have been 
studied \cite{MARC}. In this section, we start our discussion from 
following effective Lagrangian:
\begin{eqnarray}
{\cal L}_{eff}&=&
-
\frac{4 \kappa G_{F}}{\sqrt{2}}
\sum_{l=e,\mu ,\tau}
\left[
~~C_{SM}^{l}(\bar{s}_{L}\gamma^\mu d_{L})
(\bar{\nu}_{L,l}\gamma_{\mu}\nu_{L,l})
+C_{LR}^{l}(\bar{s}_{L} d_{R})(\bar{\nu}_{L,l}\nu_{R,l})
\right.
\nonumber \\
& &
\left.
\hspace{5cm}
+C_{RL}^{l}(\bar{s}_{R} d_{L})(\bar{\nu}_{R,l}\nu_{L,l})
+ h.c.
\right],
\end{eqnarray}
where $\kappa= \alpha/(2 \pi \sin^2\Theta_{W})$.
The first term is given by  SM contributions \cite{BURA,INAM} and the second 
and third terms are given by new contributions in our  model. 
The scalar and tensor operators may generally appear from box
diagram when one consider a model which contains the right handed
charged gauge boson $W_{R}$. In this paper, we only discuss the 
contribution from the scalar operator and show the explicit 
form of the coefficients $C_{LR}^l$ and $C_{RL}^l$.

First we show the decay amplitude for the neutral K meson, 
$K_{L}$ and $K_{S}$ ($ |K_{L,S}>\equiv p |K^0> \pm q 
|\bar{K}^{0}>, ~CP |K^0> \equiv - |\bar{K}^{0}>$). 
We assume neutrinos are massless, thus each term in the effective
Lagrangian do not interfere in the decay process. 
The decay amplitude $A(K_{L,S}\rightarrow\pi^0 \bar{\nu}\nu)$ are:
\begin{eqnarray}
A(K_{L,S}\rightarrow \pi^0 \bar{\nu}_{l}\nu_{l})
&=&
-\frac{G_{F}\kappa}{\sqrt{2}}
\left(
\left( p C_{SM}^{l} \mp q C_{SM}^{l \ast} \right)
<(\bar{s}d)_{V}>
\left( \bar{\nu}_{l}\nu_{l} \right)_{V-A}
\right.
\nonumber \\
&&
\left.
+
\left(p( C_{LR}^l+C_{RL}^l)\pm q(C_{LR}^l+C_{RL}^l)^{\ast}\right)
<\bar{s}d> (\bar{\nu}_{l}\nu_l)
\right.
\nonumber \\
&& 
\left.
+
\left(
p (C_{LR}^l -C_{RL}^l)\mp q (C_{LR}^l - C_{RL}^l )^{\ast}
\right)
<\bar{s}d> (\bar{\nu}_{l}\gamma_{5}\nu_{l})
\right),
\end{eqnarray}
where $<{\cal O}>$ is a short hand notation for 
$<\pi^0|{\cal O}|K^0>$.

The CP conserved limit corresponds to $p=q$ with all 
coefficients $C_{SM}^l$, $C_{LR}^l$ and $C_{RL}^l$ 
being  real. 
In this limit, the decay amplitude $A(K_L \rightarrow \pi^0 \bar{\nu}_{L}\nu_{L})$ 
through the V-A interaction is zero, while 
$A(K_S \rightarrow \pi^0 \bar{\nu}_{L}\nu_{L})$ is nonzero 
and the decays through the scalar operators $A(K_{L,S}\rightarrow 
\pi^0 \bar{\nu}_{R}\nu_{L},\bar{\nu}_{L}\nu_{R})$ 
remain non-zero generally.
In LR symmetric case ($C_{LR}=C_{RL}$), $K_S$ decay through 
the scalar operators are the CP violating mode, while CP is conserved for
$K_L$ decay. 
Thus decays of neutral K meson in LR symmetric case 
are summarized as follows:
%------------------   K_{L,S} decay tabular   -------------
\begin{eqnarray}
&&
%\hspace{-1cm}
\bullet
K_{L} \mbox{decay} 
\left\{
\begin{array}{lll}
(\bar{s}d)_{V-A} (\bar{\nu}_l\nu_l)_{V-A}&\Rightarrow&CP \hspace{-15pt}~/ \ \  ,\\
(\bar{s}d)_{S}(\bar{\nu}_l \nu_l)_{S}&\Rightarrow&CP ~\mbox{Conserving\ ,}
\end{array}
\right .
\nonumber \\
%
%
%\hspace{-1cm}
&&
\bullet
K_{S} ~\mbox{decay} 
\left\{
\begin{array}{lll}
(\bar{s}d)_{V-A} (\bar{\nu}_l\nu_l)_{V-A}&\Rightarrow&CP ~\mbox{Conserving\ ,} \\
(\bar{s}d)_{S} (\bar{\nu}_l\nu_l)_{S}&\Rightarrow&CP \hspace{-15pt}~/ \ \ . 
\end{array}
\right .
\nonumber
\end{eqnarray}
Experimentally we do not observe neutrinos, so the pion 
energy spectrum is obtained by summing these contributions 
which have different CP properties each other. 
The $K_{L}$ decay through V-A operator is  suppressed due to CP symmetry, but decays 
through the scalar operators are CP conserved ones and furthermore 
its matrix element is enhanced by use of equation of motion:
\begin{eqnarray}
<\pi^0|\bar{s}d|K^0>
&=&
\frac{p_{-}\cdot<\pi^0|(\bar{s}d)_{V-A}|K^0>
}{m_{d}-m_{s}} 
\sim  \frac{M_{K}}{m_{s}} \times M_{K} f^{\pm}, 
\end{eqnarray}
where $<(\bar{s}d)_{V-A}>=f_{+} p_{+}^{\mu}+f_{-} p_{-}^{\mu}$ and
$p_{\pm}=P_{K}\pm p_{\pi}$. 
Thus, the contribution of the scalar interaction to 
the decay amplitude $A(K_{L} \rightarrow \pi^0 \nu \bar{\nu})$
is sizable and dominates at CP conserved limit.

The decay amplitude for the charged K meson $A(K^{+} \rightarrow \pi^{+}
\nu \bar{\nu})$ is obtained in the same way:
\begin{eqnarray}
A(K^{+} \rightarrow \pi^+ \bar{\nu}_{l}\nu_{l})
&=&
-\frac{G_{F}}{\sqrt{2}}\kappa 
\left(~
 C_{SM}
<(\bar{s}d)_{V}> (\bar{\nu}_{l}\nu_{l})_{V-A},
+
\left(C_{LR}+C_{RL}\right)<\bar{s}d> (\bar{\nu}_{l}\nu_{l})
\right. 
\nonumber \\
& & \hspace{3cm}
\left.
+\left(C_{LR}-C_{RL}\right)
<\bar{s}d> (\bar{\nu}_{l}\gamma_{5}\nu_{l})~
\right)
\label{eqn:amp},
\end{eqnarray}
where $<{\cal O}>=<\pi^+|{\cal O}|K^+>$.

Now we show explicit form of the coefficient function 
$C_{LR}^l$ in LR model \cite{MORO}. We choose LR symmetric
case, $C_{LR}^l=C_{RL}^l$. 
There are two types of diagrams, penguin and box, which 
contribute to the process we are interested in.  
However, only the box diagrams produce the scalar operator
$(\bar{s}d)_{S}(\bar{\nu}\nu)_{S}$ in the effective Lagrangian. 
Thus we do not consider the contributions from penguin diagrams 
in this paper. There are tensor operators $(\bar{s}\sigma^{\mu\nu}d)
(\bar{\nu}\sigma_{\mu\nu}\nu)$ from box diagram in LR model, 
which will be discussed in the other place \cite{PREP}. 
 
We calculate box diagrams, in which left handed $W_{L}$ boson (W in SM) and 
right handed gauge boson $W_R$ are exchanged [Fig.1]. 
There are corresponding charged Higgs diagrams due to the gauge 
invariance. The internal upper fermion lines correspond to the ordinary
and singlet quarks, the lower correspond to SM and singlet leptons.

\begin{figure}[h]
\begin{center}
\leavevmode\psfig{file=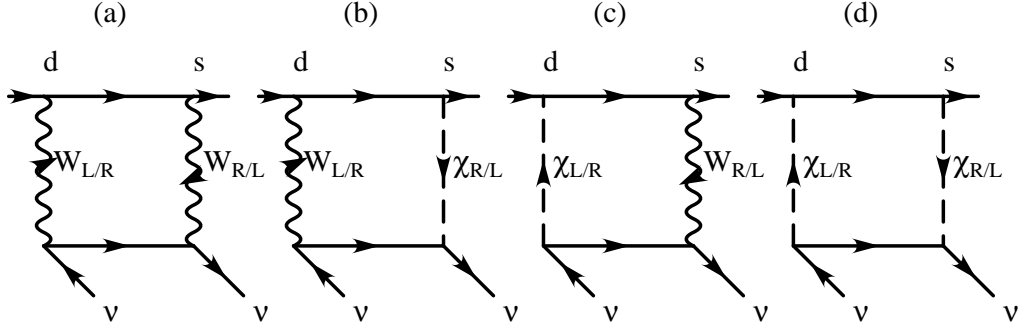,width=14cm}
\caption{Box diagrams which contribute the effective 
Lagrangian for the process
$K \rightarrow \pi \nu \bar{\nu}$. 
(a) is a contribution from $W_{L}$ and $W_{R}$. (b) and (c) are gauge
boson and unphysical Higgs contributions. (d) is a contribution from 
unphysical Higgs $\chi_{L}$ and $\chi_{R}$.}
\end{center}
\end{figure}

The coefficients in the effective Lagrangian are:
\begin{eqnarray}
C_{LR}^l=C_{RL}^l
&=&
\cos\theta_{L}^{s}\cos\theta_{R}^{d}\sum_{q=u,c,t}
(V^{\ast~qs} V^{qd}) \cos\theta_{L}^{q} \cos\theta_{R}^{q} 
\nonumber \\
&&
\times
\beta \sqrt{x_q y_l}
\left[ d(x_q,y_l) -d(X_q,y_l) -d(x_q,Y_l) +d(X_q,Y_l) \right],
\nonumber 
\end{eqnarray}
where $\theta_{L/R}^{q}$ is a mixing angle between singlet left/right
handed quark and corresponding doublet quark, $V^{ij}$ corresponds to 
$3\times 3$ CKM matrix element. $x_{q}, X_{q}, y_{l},Y_{l}$ and 
$\beta$ are dimensionless parameters defined by:
\begin{equation}
x_{q}=\frac{m_{q}^{2}}{M_{W}^{2}}, \hspace{1cm}
X_{q}=\frac{M_{q}^{2}}{M_{W}^{2}}, \hspace{1cm}
y_{l}=\frac{m_{l}^{2}}{M_{W}^{2}}, \hspace{1cm}
Y_{l}=\frac{M_{l}^{2}}{M_{W}^{2}}, \hspace{1cm}
\beta=\frac{M_{W}^{2}}{M_{R}^{2}},
\end{equation}
where $m_{q/l}$ and $M_{q/l}$ is a mass of the ordinary quark/lepton
and mass of the heavy singlet quark/lepton. 
The function $d(x,y)$ is defined by:
\begin{eqnarray}
d(x,y)
&=&
\frac{4+\beta x y}{4}
\left[\frac{ x \ln x }{ (1-x)(x-y)(1-\beta x) }+( x \leftrightarrow y)-
\frac{\beta \ln \beta }{ (1-\beta x)(1-\beta y)(1-\beta)} \right]
\nonumber \\
&&
-
\frac{1+\beta}{4}
\left[
\frac{x^{2}\ln x 
     }{ (1-x)(1-\beta x)(x-y) }
+(i \leftrightarrow j)
-\frac{\ln \beta 
      }{(1-\beta)(1-\beta x)(1-\beta y)}
\right] .
\end{eqnarray}

The coefficient $C_{RL}^l$ ($l=e, \mu$) for 
electron and muon is negligibly small and only $C_{LR}^{\tau}$ contributes to 
the process significantly. 
The dependence on the internal quark is very similar to the one in the SM.  
For the top sector, coefficients are large by the top quark mass, 
but suppressed by the CKM factor $V^{ts \ast}V^{td} \sim \lambda^5$
compared to the one for the charm sector. 
The coefficient for the up quark sector is negligible. 
All the coefficients of the scalar interactions are suppressed 
by the $\beta$, so the V-A interaction in SM dominates when 
$M_{R}$ becomes large. 
 
%------------------------ Sec. 4 ---------------------------------
\section{Pion Energy Spectrum}
In this section, we present the pion energy spectrum by using the coefficient 
in section 2.
In the process $K^+ \rightarrow \pi^{+} \nu \bar{\nu}$, contribution
from the scalar interactions are tiny compared to the one in  SM. 
The pion energy spectrum for the decay $K^{+} \rightarrow \pi^+ \nu
\bar{\nu}$ is:
\begin{eqnarray}
\frac{dB^{K^{+} \rightarrow \pi^+ \nu_{l} \bar{\nu}_{l}}}{d x_{\pi}}
&\sim&
\sqrt{x_{\pi}^2-4 \delta^2}
\left[ (x_{\pi}^2 -4 \delta^2)| C_{SM}^l|^2
+
3 \hat{t} \left(\frac{M_{K}}{m_{d}-m_{s}}\right)^2 
\left(1-\delta^2+\xi ~\hat{t} \right)^{2}
| C_{LR}^l|^2 
\right] , 
\end{eqnarray}	
%%%%%%%%%%%%%%%%%%%%%%%%%%%%%%%%%%%%%%%%%%%%%%%%%%%%%%%%%%%%%%%%%%%%%%%%%%%
%\begin{eqnarray}
%\frac{d\Gamma}{d x_{\pi}}(K^{+} \rightarrow \pi^+ \nu_{l} \bar{\nu}_{l})
%&=&
%\frac{ M_{k}^5  G_F^2 \kappa^2 (f_{+}^{K^+ \pi^0}(q^2))^2 
%}{12 (2 \pi)^3}
%\sqrt{x_{\pi}^2-4 \delta^2}
%\nonumber \\
%&& \hspace{-4cm}\times 
%\left[ (x_{\pi}^2 -4 \delta^2)| C_{SM}|^2
%+
%3 \hat{t} \left(\frac{M_{K}}{m_{d}-m_{s}}\right)^2 
%\left(1-\delta^2+\xi ~\hat{t} \right)^{2}
%| C_{LR}|^2 
%\right]
%\end{eqnarray}	
%%%%%%%%%%%%%%%%%%%%%%%%%%%%%%%%%%%%%%%%%%%%%%%%%%%%%%%%%%%%%%%%%%%%%%%%%%%
%
where $x_{\pi}$ is a normalized pion energy defined by $x_{\pi}=2 
E_{\pi}/M_{K}$, and  $\delta=m_{\pi}/M_{K}$, $\xi=f_{-}^{K^+\pi^0}/f_{+}^{K^+\pi^0}$ 
and $\hat{t}=(1+\delta^2-x_{\pi})$ are defined.

The process
$K_{L} \rightarrow \pi^0 \nu \bar{\nu}$ is more sensitive to probe the 
scalar interactions as discussed in the previous section. 
The energy spectrum is given by:
\begin{eqnarray}
\frac{dB^{K_{L/S} \rightarrow \pi^0 \nu_{l} \bar{\nu}_{l}}}{d x_{\pi}}
&\sim&
\sqrt{x_{\pi}^2 -4 \delta^2}
\left[ 
(x_{\pi}^2 -4 \delta^2)
|p C_{SM}^l \mp q C_{SM}^{l \ast}|^2
\right.
\nonumber \\
&&
\left.
+ 
3 \hat{t}\left(\frac{M_{K}}{m_{d}-m_{s}}\right)^2 
\left(1-\delta^2+\xi ~\hat{t}\right)^2 
|p C_{LR}^l \pm q C_{RL}^{l \ast} |^2 
\right] .
\end{eqnarray}
%
%%%%%%%%%%%%%%%%%%%%%%%%%%%%%%%%%%%%%%%%%%%%%%%%%%%%%%%%%%%%%%%%%%%%%%%%%%%
%\begin{eqnarray}
%\frac{d\Gamma}{d x_{\pi}}(K_{L/S} \rightarrow \pi^0 \nu_{l} \bar{\nu}_{l})
%&=&
%\frac{M_K^5 G_F^2 \kappa^2 (f_{+}^{K^+ \pi^0}(q^2))^2 }{24 (2\pi)^3} 
%\sqrt{x_{\pi}^2 -4 \delta^2}
%\nonumber \\
%&& \hspace{-4cm}\times
%\left[ 
%(x_{\pi}^2 -4 \delta^2)
%|p C_{SM} \mp q C_{SM}^\ast|^2
%+ 
%3 \hat{t}\left(\frac{M_{K}}{m_{d}-m_{s}}\right)^2 
%\left(1-\delta^2+\xi ~\hat{t}\right)^2 
%|p C_{LR} \pm q C_{RL}^\ast |^2 
%\right],
%\end{eqnarray}
%%%%%%%%%%%%%%%%%%%%%%%%%%%%%%%%%%%%%%%%%%%%%%%%%%%%%%%%%%%%%%%%%%%%%%%%%%%
%
The first term is the SM contribution from V-A operator 
$(\bar{s}d)_{V-A}$, and the second is the contribution from the scalar operator 
which has a enhancement factor $ M_{K}^2/(m_{d}-m_{s})^2$ 
 in the matrix element. 
Form factors are related to the one of experimentally well 
known leading decay mode $K^+ \rightarrow \pi^0 e^+\nu$.

We show the pion energy spectrum $\frac{d B(x_{\pi})}{d x_{\pi}}$,
in which we summed over neutrino flavors and normalized by a factor $10^{-10}$,  in 
Fig.2. %
\vspace{-1cm}
\begin{figure}[H]
\begin{center}
\leavevmode\psfig{file=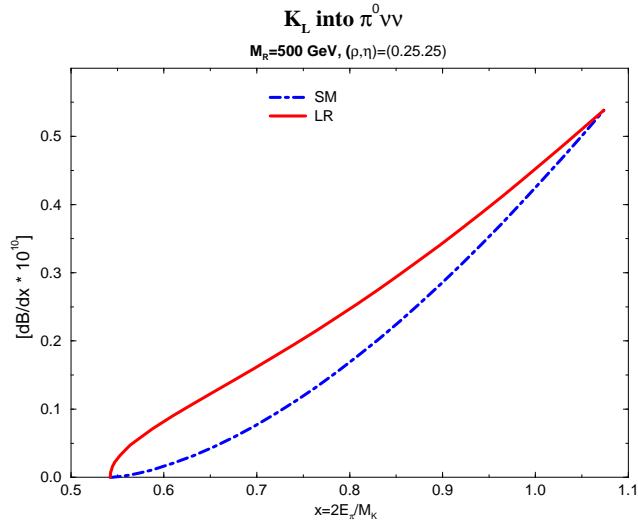,width=10cm,angle=-90}
\caption{
The solid line is a pion energy spectrum in LR model, 
doted is the SM prediction, where  we take  the right handed gauge 
boson mass $M_{R}=500 GeV$ and $\rho=\eta=0.25$. 
LR contribution is large in the low energy region 
$x_{\pi} \sim 2 \delta$, while in the high energy 
region the SM contribution dominates. 
}
\end{center}
\end{figure}
%

%--------------------------------------------
\section{Conclusion}
We have discussed rare decays of K mesons, 
$K^+ \rightarrow \pi^+ \nu \bar{\nu}$ and 
$K_{L,S} \rightarrow \pi^0 \nu \bar{\nu}$ including 
the contributions from scalar operators in the effective 
Lagrangian, which is produced from LR box diagram,
For the decay  $K^{+}\rightarrow \pi^+ \nu \bar{\nu}$, 
contribution from LR model is small compared to the one of SM.  
For the decay  $K_{L}\rightarrow \pi^0 \nu \bar{\nu}$, there is  
significant contribution from the scalar operator, especially in the low energy 
region of the pion energy spectrum, which amount to about 30 \%
enhancement to total branching ratio, with the parameters 
$\rho,\eta=0.25$ and right handed gauge boson mass $M_{R}=500$ [GeV]. 
Thus, measuring the decay $K_{L}\rightarrow \pi^0 \nu \bar{\nu}$ 
precisely is very important to probe the effect from new physics.
\section*{\bf Acknowledgments} 
This work is supported by the
Grant-in-Aid for Joint Internatinal Scientific Research
($\sharp 08044089$,  Origin of CP and T violation and flavor
physics) and by Grant-in-Aid for
Scientific Research on Priority Areas (Physics of CP violation)
from the Ministry of Education, Science and Culture, Japan.
%%%%%%%%%%%%%%%%%%%%%%%%%%%%%%%

%%%%%%%%%%%%%%%%%%%%%%%%%%%%%%%
\end{document}